\documentclass[a4paper]{VisionStyle}
\usepackage{epsfig}
\def \xmm {XMM-Newton } 

\begin{document}
\title{EPIC in orbit background}

\author{D.H.\,Lumb\inst{1} } 

\institute{ESA Payload Technology Division, Research and Scientific Support Department, ESTEC, Postbus 299, NL-2200 AG Noordwijk, The Netherlands}

\maketitle 

\begin{abstract}
We briefly describe the methods used in compiling a set of high galactic latitude background data. The characteristics and limitations of the data which affect their use as a template for analysing extended objects are described. We briefly describe a spectral fitting analysis of the data which reveals a normalisation for the extragalactic background  of 11.4 keV cm$^{-2}$ s$^{-1}$ sr$^{-1}$, which implies the fraction of hard X-ray background presently resolved by various Chandra observations is towards the lower limit of their estimates.

\end{abstract}

\section{introduction}\label{dlumb-WA2b_sec:reqs}
Many users have asked for background data sets with which to analyse extended
objects, particularly where determination of background is difficult from their own data sets. We have developed some event lists for each EPIC camera that 
could be used {\em ad interim}. These files have an extension containing 
a calibrated event list in the same format as produced by SAS. Compared with
previous efforts, the new sets (ca. December 2001) are significantly longer to allow improved signal:noise (typically
the exposure duration is about an order of magnitude longer than the normal GO 
data set); the  data are homogeneously collected from high galactic fields with no bright sources; the data set uses homogenously the THIN filter only and
{\rm EXPOSURE,STDGTI, BADPIX} and {\rm EVENTS} extensions of the files have all been 
concatenated and made coherent, so that in principle the observer could use
them as a coherent data set for extractions via various $SAS$ or $ftool$
procedures, without making the manual corrections and calculations that were
required for the previous incarnation.

Many diverse aspects of the XMM-Newton radiation environment were
discussed extensively at a Workshop held in December 2000.
The reader is invited to look at the links under {\rm /docs/documents/} pages
of the VILSPA web site, 
particularly the document: {\rm CAL-MIN-0002-0-0.html}. This has pointers to viewgraphs and minutes
from the Workshop.
The present note concerns {\em quiescent} background (i.e. after removal of proton flares via. GTI filtering).
\section{Data Selection} 
\subsection{Field Locations} 
We have had to resort to compiling 
data from blank sky fields. To do so with realistic S:N ratios 
for each of 4 instrument modes would impose an unacceptable penalty 
on Guest Observer science programme time, so we concentrated on  
the Full Frame imaging modes which are generally used for the faint extended 
objects. To minimise 
statistical uncertainties, the effective exposure duration in our data sets  
should be an order of magnitude longer than that of the typical Guest Observer  
exposure. 
Again it was an unrealistic proposition to observe 
with the 3 EPIC optical blocking filters for $\sim$300ks each, specifically 
 to obtain this required data, therefore we decided to make use of a variety 
of Guaranteed and Calibration Time observations of ``blank'' fields to compile 
our data serendipitously.
 
Suitable fields  
were almost all taken with the filter in THIN position, although this imposes additional complications for those users 
who might need to use a thicker optical blocking filter.
Nevertheless as the soft diffuse X-ray component is more spatially and  
spectrally variable than the harder X-ray emission components this was not the  
major driver for our analysis.

Table 1 summarises the location of the fields selected. 

  \begin{table*}[bht]
  \caption{Summary of target locations compiled}
  \label{dlumb-WA2b_tab:tab1}
  \begin{center}
    \leavevmode
    \footnotesize
\begin{tabular}{l l l c c l l} 
      \hline \\[-5pt]
   RA &Dec &Date of &Duration &N$_{H}$ &L II& B II\\   [+5pt]
   &     &observation&(ks) &(10$^{20}$cm$^{-2}$)& & \\ [+5pt]
      \hline \\[-5pt]     

02:18:00&-05:00:00&2000-07-31&60&2.5&169.7&-59.8\\ 
02:19:36&-05:00:00&2000-08-04&60&2.55&170.35&-59.5\\ 
02:25:20&-05:10:00&2001-07-03&25&2.7&172.3&-58.6\\ 
02:28:00&-05:10:00&2001-07-06&25&2.7&173.5&-58,2\\ 
10:52:44&+57:28:59&2000-04-29&70&0.56&149.3&53.1\\ 
12:36:57&+62:13:30&2001-06-01&90&1.5&125.9&54.8\\ 
13:34:37&+37:54:44&2001-06-23&80&0.83&85.6&75.9\\ 
22:15:31&-17:44:05&2000-11-18&55&2.3&39.3&-52.9\\ 
\end{tabular} 
  \end{center}
\end{table*}

 \subsection{Data Generation} 
The data sets were processed using the pipeline processing of XMM-Newton Science  
Analysis Subsystem 5.2, in order to generate calibrated event lists for each 
EPIC camera.  Proton flare GTI filtering was made on time bins of $\sim$100s at count rates $\geq$ 45 (20) events/bin in PN (MOS). 
Next we removed the signature of bright field sources by forming images in the 0.5 - 2keV band and running SAS task $EBOXDETECT$. Approximately 10 objects per field were identified 
and an exclusion of 25 arcsec radius applied around the detected centroids. 
This exclusion cannot be directly associated with a source flux level, but  
individual analysis of the fields confirmed the estimates  
based on LogN-LogS curves (\cite{dlumb-WA2b:hasing}), that sources of flux 
brighter than about 1 - 2 $\times$ 10$^{-14}$ ergs cm$^{-2}$ s$^{-1}$ (0.5-2 
keV) have had about 80\% of their flux removed with this recipe. 
Finally the screened data were co-added, and the exposure and GTI extensions 
of the FITS data files carefully added and concatenated together.

\section{Data Characteristics} 
\subsection{Fluorescent Emission Lines} 
The passage of charged particles through the cameras is associated with 
generation of fluorescent X-ray emission. This emission is most clearly seen 
in the form of emission line energies characteristic of the camera body materials 
(aluminium and stainless steel components for example). The construction of 
the two EPIC camera types is quite different, leading to substantially 
different manifestations of these features. 
The outer 6 of 7 CCDs detect more Al K radiation due to their closer proximity 
to the aluminium camera housing. Si K emission however is concentrated 
along the edges of some CCDs. 
In contrast the PN camera has a relatively intense contribution from  
energies around the Cu K line. What is 
notable is that 
this emission is spatially variable (\cite{dlumb-WA2b:frey}). 

The consequences for ignoring the spatial variation of these background 
features could be dramatic. XMM-Newton is possibly the observatory 
of choice for spectrally-resolved imaging of large clusters, for example to 
map radial temperature and element distributions. However the variable Al 
and Si background lines would compromise abundance determinations of  
cluster emission lines with moderate redshifts, while the variable high energy 
background would bias temperature measurements at large radii. These  
difficulties  
should be alleviated if the proposed background templates prove to be  
representative.     
\subsection{Unrejected Particle Background} 
A background component that is relatively constant in spectrum and with  
position, is 
the remnant events generated by charged particles which are not rejected by 
on-board or ground processing. Compared with a maximum predicted primary  
cosmic ray particle rate of 4 cm$^{-2}$ s$^{-1}$, the measured rates in all the  
CCD cameras on \xmm  since launch have lain in the range 2 - 2.5  
cm$^{-2}$ s$^{-1}$.  
The internal spectrum in the MOS1 camera, below 10keV, after selection for X-ray event characteristics (the SAS 
attributes {\em (\#XMMEA\_EM\&\&PATTERN in [0:12])} ) is 
0.026 events cm$^{-2}$s$^{-1}$. Of this 0.021 ($\pm$0.0022) cm$^{-2}$s$^{-1}$ is in 
the flat spectrum component, the remainder in emission line components and 
noise component increasing to lower energies. The flat spectrum count rate 
  implies a cosmic ray rejection efficiency of $\sim$99\%  as expected.  
The equivalent spectrum in the PN cameras after 
applying the SAS attributes {\em \#XMMEA\_EP \&\& PATTERN in [0:4]} is 0.039 cm$^{-2}$s$^{-1}$ is in 
the flat spectrum component, and 0.034 cm$^{-2}$s$^{-1}$ in the fluorescent
lines.  This could imply a reduced
rejection efficiency than MOS or a higher detection efficiency to local $\gamma$'s as a 
consequence of the much thicker pn detection thickness.
\subsection{Low Energy Artefacts} 
The most obvious features in the lowest energy band are effects of ``bad  
pixels''. A pixel which is consistently bad is flagged for removal  
on-board by loading a table of positions to be blanked out. 
Some pixels ``flicker'' on and off  (\cite{dlumb-WA2b:hop})with low  
recurrence rate ($\leq$1\%), and  
the efficiency for finding them post-facto in the SAS pipeline is dependent on  
many 
factors, so that some such events may occur in the background template and not  
the observer's data set and vice versa.  
 
In the PN camera there are occasional blocks of bright pixels, typically 
4 pixels in height. Their presence varies from observation to observation. 
They arise from an artefact of the CCD offset bias level calculation at the  
start of  
each observation. 

\section{Spectral Analysis} 
\subsection{Internal Background} 
\begin{figure*}[ht]
  \begin{center}
    \epsfig{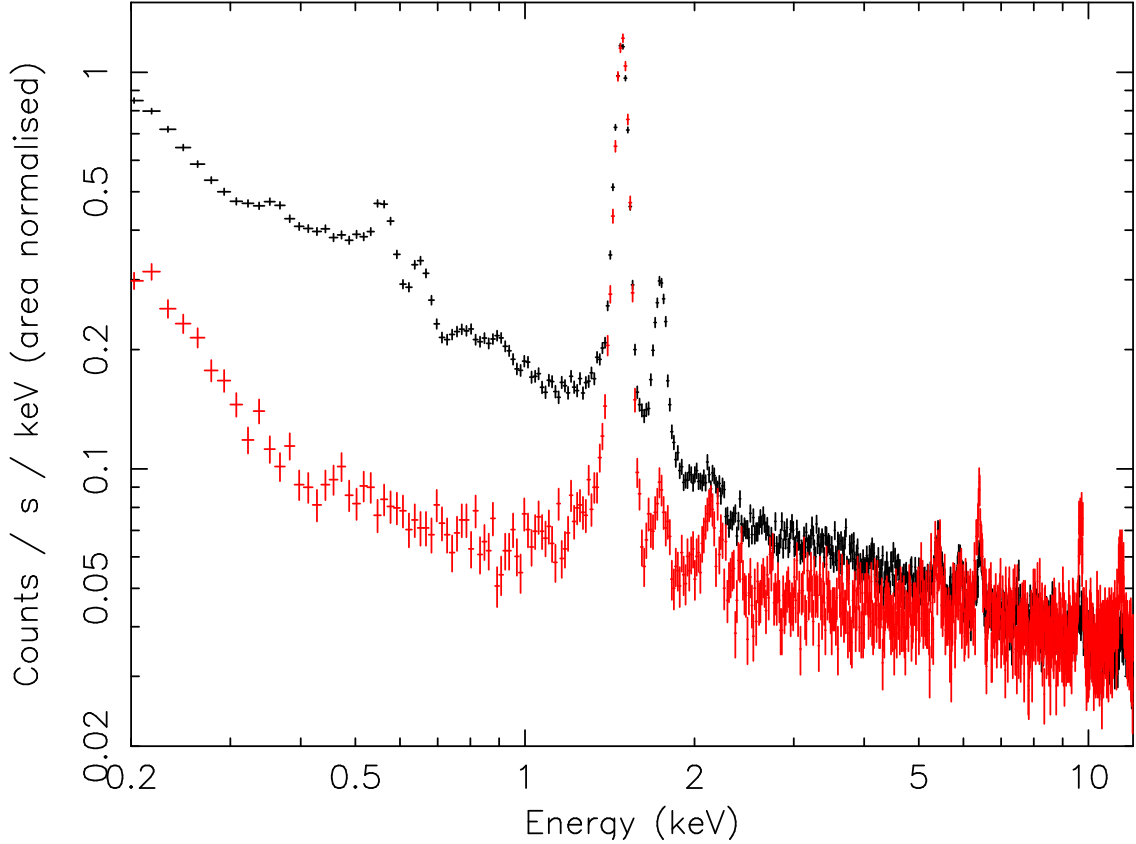}
  \end{center}
\caption{ Comparison of internal (red) and total (including cosmic diffuse) background spectra in MOS cameras. }  
\label{dlumb-WA2b_fig:cxbint}
\end{figure*}

In principle one can use data from outside the filter rim as a measure of the true
internal background. However as noted, this is not strictly true due to the
spatial variation in fluorescent emission.
 This was easiest to perform in MOS, and therefore we  modeled the  
internal background, with  multiple Gaussian functions to characterise  
these emission lines, imposed upon a broken power-law to describe the continuum  
due to unrejected particle backgrounds, and low energy noise. 
  In Fig~\ref{dlumb-WA2b_fig:cxbint} we show the comparison of this internal background component  
with the total spectrum including the CXB. The internal component should be scaled  
 with its particle continuum according to the ratio of in- and out-of-field 
detector areas, but the emission line intensities must be allowed to vary (in order 
to account for the aforementioned spatial variation). 

 At energies $\leq$5keV the CXB 
dominates, with a clear signature of low energy emission lines at E$\leq$1 keV 
not present in the internal background, but most likely from a thermal cosmic 
plasma spectrum.   

\subsection{Cosmic X-ray Background} 
In the case of the MOS data files, where we are more secure of a subtraction of the internal background component, 
we attempted a simple spectral fit to the diffuse background spectrum. We adopted an empirical two temperature MEKAL (Mewe et al  \cite{dlumb-WA2b:mewe})  
plus power-law component to describe 
the cosmic diffuse X-ray background. The former represent the contribution from the supposed Galactic Halo, and the latter the extragalactic unresolved AGN  
population.  Temperatures of $\sim$0.08keV and $\sim$0.21keV together with a
power law 1.4 could be recovered.
We made an estimate of the sub-keV emission in some of the different fields noting differences in the measured temperature and flux. This variability in emission and lack of detailed temperature measurement capability hampers a more accurate determination of Galactic background properties.  For example the measured
mean 
deviation from field to field of 2 - 10 keV flux is about 3.5\%, consistent with a uniform diffuse extragalactic background. On the other hand the mean deviation
of 0.2 - 1 keV flux is about 35\% from field to field. 
 
The recovered spectral fitting parameters are consistent with a number of
investigations of the diffuse X-ray background, which leads us to believe that
the user should be able to use the sets with suitable caveats to analyse more
general conditions with different filters and Galactic latitude, as described below. The calculated normalisation of extragalactic background was 11.4 keV cm$^{-2}$ s$^{-1}$ kev$^{-1}$ sr$^{-1}$, which is comparable with determination by SAX, but
places the fraction of {\rm Chandra}-resolved harder X-ray background towards the
lower limits of claimed fractions (70\% rather 95\%).
\section{Caveats for Use} \label{sec:caveat} 
The EPIC cameras allow a choice of optical blocking filter to prevent  
contamination by optically bright targets. For most extended and/or faint  
extragalactic targets, such contamination is negligible, and the thinnest filter  
can be employed, as indeed was the case for all the observations compiled in  
this work. However should the observer have chosen a thicker filter, the  
transmission of diffuse background X-rays (and no doubt any remnant proton flux)  
will be reduced. Therefore the observer would have to model the differences in  
soft component based on the response matrices and limited knowledge of the  
expected Galactic emission. 
 
The observer must take care to extract background from an appropriate location  
in the field of view. Most simply the extraction region should be defined in  
detector co-ordinates  to match closely the region of the desired science  
target. In some applications the user would subtract data in sky co-ordinates.  
In this case the template event lists could be recast to mimic the nominal  
pointing direction of the observer's field (for example using SAS task  
$ATTCALC$ invoking the attributes $attitudelabel=fixed$  $withatthkset=N$  $refpointlabel=user$ )
 
We noted that despite the standard recipe for filtering proton flares, the  
background rate as measured in 1-10keV bands exhibited some remnant flares. This  
implies that at lower energies, the proton flares turn on more slowly, yet  
before the main flare component. We have not chosen to further force stringent data cuts in the background template files, so that the general observer has more leeway to make an additional  
selection of the template files to match his/her own data selection.  We  
emphasise that in analysing the spectrum of diffuse X-rays in the field, the  
recovered spectral slopes steepen with more thorough flare mitigation. Careful  
comparison of the recipes used for GTI creation must be made. Nevertheless it is  
possible that the lowest level proton fluxes are spectrally variable, and no  
complete subtraction can be made.   
  
Although we noted that point sources have been removed or significantly diluted,  
careful examination of the images derived from these event lists reveals  
intensity fluctuations. Over scales of arcminutes appropriate to extended  
sources, it is not expected to be a significant problem, and indeed  
{\em representative} of unresolved background. However if the user should try to  
extract spectra from regions comparable with the mirror point spread function  
scale,  
then manual inspection might be necessary to guard against a local deficit or  
excess of counts arising from treatment of point sources in the template files. 
 
As noted previously there are particular defects to be expected in the lowest  
energy spectral ranges. Furthermore, at the time of writing, the calibration of  
the EPIC soft X-ray spectral response awaits completion. The transmission of  
filters at energies $\leq$250eV is difficult to measure, the CCDs' calibration  
at the ground synchrotron facility was not performed at energies $\leq$150eV and  
the detailed redistribution of signal from photons of energies $\sim$1keV into  
partially collected  events in the softest band was also not determined  
completely in ground measurements. For the time being the extension of spectral  
analysis to {\em any} data below 250eV should be treated with caution. 
 
For the highest fidelity determination of background appropriate to the 
observer's own data the compensation for galactic soft X-ray component, changing
cosmic ray rates and different filters must be made. It is intended 
to provide tools within the \xmm SAS environment to achieve this, but most 
steps should be achievable manually. 

\begin{itemize} 
\item Following suitable flare screening, define a background region (B) and  
extract the observed spectrum (C$_{back}$) from the observer's data set. From
an identical region in the template file the observed spectrum (T$_{back}$)
should provide a measure of variability of CR component by checking count rates for
E$\geq$5keV and/or the fluorescent emission line normalisations.
\item  To estimate a better  internal background spectrum for the observer's data set (C$_{inst}$),
determine a predicted cosmic background spectrum for the observer's region based on 
ROSAT ASS maps, hydrogen column etc.. An experimental tool is available at HEASARC web site.
Create response matrices for the background region (here is where you can
introduce the effect of different filters). 
Fold this cosmic spectrum through the response matrices to obtain a 
{\em predicted} cosmic component  for the background region, (C$_{cos}$). 
C$_{inst}$ = C$_{back}$ - C$_{cos}$
\item A similar approach with the template files showed that with a weighted average N$_{H}$
of 1.7 10$^{20}$ leads to a R$_{45}$ PSPC count rate of 1.3s$^{-1}$ in 144arcmin$^{2}$ and a
0.47-1.21 flux of 1.67 10$^{-11}$ ergs cm$^{-2}$ s$^{-1}$ for a 0.2keV R-S spectrum. This
could likewise be used to make an estimate of the internal background of the template 
file region in order to better estimate the scaling factor (K) for CR component.
T$_{inst}$ = T$_{back}$ - T$_{cos}$ and K $\sim$C$_{inst}$/T$_{inst}$
\item Repeating the same exercise for the source region in both template and observed
data sets could lead to a {\em predicted} background data spectrum comprising the
scaled internal component, and the predicted galactic component with the appropriate 
filter responses.
\end{itemize}

\end{document}